\documentclass[runningheads]{llncs}

\usepackage{makeidx}  
\usepackage[clock]{ifsym}

\usepackage{graphicx}
\usepackage{amsmath}
\usepackage{mymacros}
\usepackage{amssymb}
\usepackage{graphicx}
\usepackage[table,xcdraw]{xcolor}
\usepackage{wrapfig}
\usepackage[makeroom]{cancel}
\usepackage{cite}
\usepackage{enumerate}
\usepackage{subcaption}

\usepackage{times}
\usepackage{adjustbox}
\usepackage[ruled,vlined,linesnumbered]{algorithm2e}
\usepackage{amsmath,algpseudocode}

\makeatletter
\renewcommand{\Function}[2]{%
  \csname ALG@cmd@\ALG@L @Function\endcsname{#1}{#2}%
  \def\jayden@currentfunction{#1}%
}
\newcommand{\funclabel}[1]{%
  \@bsphack
  \protected@write\@auxout{}{%
    \string\newlabel{#1}{{\jayden@currentfunction}{\thepage}}%
  }%
  \@esphack
}
\makeatother
\algdef{SE}{Begin}{End}{\textbf{begin}}{\textbf{end}}

\usepackage[colorlinks=true,linkcolor=blue,citecolor=blue,urlcolor=blue]{hyperref}
\usepackage[section]{placeins}
\usepackage{verbatim}
\usepackage{mathabx}
\usepackage{multirow}
\usepackage{capt-of}
\usepackage{tikz}
\usepackage{xcolor}

\usepackage{nameref}
\newcounter{mylabelcounter}

\makeatletter
\newcommand{\labelText}[2]{%
#1\refstepcounter{mylabelcounter}%
\immediate\write\@auxout{%
  \string\newlabel{#2}{{1}{\thepage}{{\unexpanded{#1}}}{mylabelcounter.\number\value{mylabelcounter}}{}}%
}%
}

\algnewcommand\algorithmicforeach{\textbf{for each}}
\algdef{S}[FOR]{ForEach}[1]{\algorithmicforeach\ #1\algorithmicdo}
\def\univs{\mathbb{U}}
\newcommand{\univ}[1]{\univs_{\mathit{#1}}}

\SetKwInput{KwData}{Input}
\SetKwInput{KwResult}{Output}

\usepackage{xcolor}
\SetCommentSty{mycommfont}

\usepackage{xcolor}
\definecolor{darkred}{rgb}{0.6, 0, 0} 
\definecolor{darkblue}{rgb}{0, 0, 0.6} 

\usepackage{manfnt}
\usepackage{tikz}
\usepackage{stackrel}
\usepackage[bottom]{footmisc}

\newcommand{\approachname}{scope-enriched }
\newcommand{\Approachname}{Scope-enriched }

\begin{document}

\mainmatter              
\title{Enriching Object-Centric Event Data with Process Scopes: A Framework for Aggregation and Analysis}


\author{Shahrzad Khayatbashi\inst{1} 
    \and Majid Rafiei \inst{2}
    \and Jiayuan Chen \inst{2}
    \and Timotheus Kampik \inst{2}
    \and Gregor Berg \inst{2}
    \and  Amin Jalali \inst{3}
    }

\titlerunning{Enriching OCED with Process Scopes: A Framework for Aggregation and Analysis}

\authorrunning{Shahrzad Khayatbashi et al.}

\institute{Linköping University, Linköping, Sweden,\\
    \email{shahrzad.khayatbashi@liu.se}
\and SAP Signavio, Berlin, Germany,\\
    \email{(firstname.lastname)@sap.com}
\and Stockholm University, Stockholm, Sweden,\\
    \email{aj@dsv.su.se}
}

\maketitle

\begin{abstract}    
Object-Centric Process Mining enables the analysis of complex operational behavior by capturing interactions among multiple business objects (e.g., orders, items, deliveries). These interactions are recorded using Object-Centric Event Data~(OCED) formats, such as the Object-Centric Event Log~(OCEL). However, existing formats lack explicit definitions of process scopes, which restricts analysis to individual processes and limits insights to a low level of granularity.
In practice, OCED often spans multiple interrelated processes, as shared objects connect events across organizational functions. This structure reflects how value is created along the organizational value chain, but introduces challenges for interpretation when process boundaries are not clearly defined. Moreover, process definitions are typically subjective and context-dependent; they vary across organizations, roles, and analytical goals, and cannot always be discovered automatically.
To address these challenges, we propose a method for embedding analyst-defined process scopes into OCEL. This enables the structured representation of multiple coexisting processes, supports the aggregation of event data across scopes, and facilitates analysis at varying levels of abstraction. We demonstrate the applicability of our approach using a publicly available OCEL log and provide supporting tools for scope definition and analysis.
\end{abstract}

\section{Introduction}\label{sec:Introduction}

Object-Centric Process Mining (OCPM)~\cite{van2019object} has emerged as a promising paradigm for analyzing complex, real-world business processes~\cite{khayatbashi2025ai, berti2023analyzing, kretzschmann2024overstock, park2023analyzing}. While traditional process mining techniques support the discovery, monitoring, and improvement of business processes, they typically rely on the assumption that each event belongs to a single case, such as an order, claim, or ticket. This single-case perspective works well for simple scenarios but often struggles in more realistic contexts, creating challenges such as artificial divergence or convergence of paths, missed interactions, and gaps in understanding how different objects interact across multiple cases~\cite{van2019object,van2020academic}.

\begin{figure}[b!]
    \centering
    \includegraphics[width=1\linewidth]{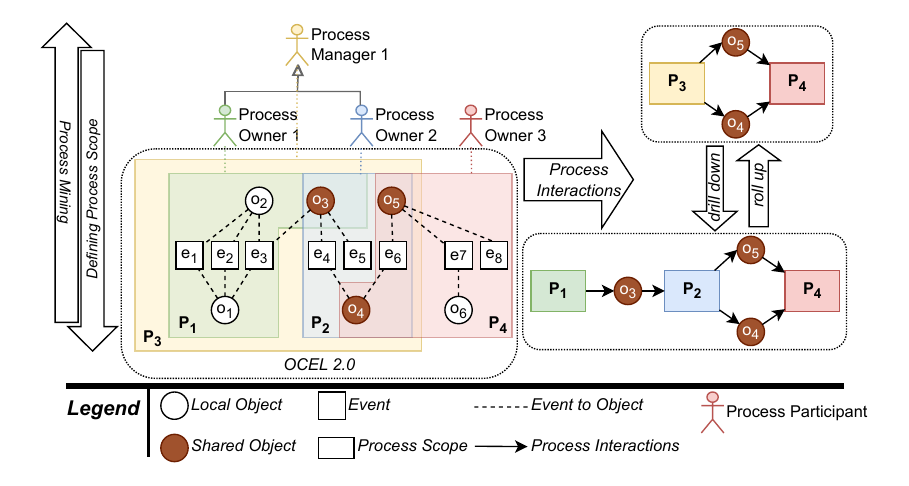}
    \caption{Example of an Object-Centric Event Log (OCEL) where events connect to multiple objects and are grouped into explicit process scopes (P1–P4). The figure illustrates how shared objects can link processes, enabling the discovery of process-level interactions.}
    \label{fig:running-example}
	\vspace{-1.2\baselineskip}    
\end{figure}

To overcome these limitations, several Object-Centric Event Data (OCED) formats have been introduced~\cite{fahland2024towards, esser2021multi, khayatbashi2024transforming}. Among them, the Object-Centric Event Log (OCEL)\cite{berti2023ocelspecification} is the most widely adopted and forms the foundation for various process mining algorithms, libraries, and tools\cite{van2020discovering, liss2025object, adams2021precision, khayatbashi2025olap, jalali2022object, adams2022ocpa}. OCEL supports events that are related to multiple business objects, enabling a more expressive and accurate representation of process behavior. For example, in a sales scenario, a single event might reference an order, a customer, and several items. \figurename~\ref{fig:running-example} illustrates an abstract OCEL where events are connected to one or more objects. This multi-object perspective better reflects how work is executed in practice and allows analysts to examine behavior from different viewpoints, revealing hidden dependencies and overlaps across processes~\cite{khayatbashi2025ai, berti2023analyzing, kretzschmann2024overstock, park2023analyzing}.

Despite its practically expressive nature, OCEL lacks explicit support for defining process scopes, which are essential for structured and context-sensitive analysis. In Business Process Management (BPM), identifying and organizing processes is a foundational step, often referred to as process identification or process architecture~\cite{dumas2018fundamentals}. Although OCED can be extracted directly from operational systems, defining the boundaries of processes requires input from domain experts such as analysts, process owners, or managers, who rely on organizational objectives, responsibilities, and analytical goals.

\figurename~\ref{fig:running-example} illustrates how different stakeholders may define overlapping or nested scopes based on their roles and perspectives. These definitions are inherently subjective yet vital for making sense of complex event data. In the absence of explicit scoping in the log, analysts typically apply manual filtering to isolate relevant subsets. While helpful, this approach is ad hoc and limits the ability to systematically explore how processes interact, overlap, or align with organizational structures.

To overcome this limitation, we propose a method for embedding explicit, analyst-defined process scopes into OCEL while preserving full compatibility with the OCEL 2.0 standard~\cite{berti2023ocelspecification}. This enables a structured representation of multiple coexisting processes within the same log. Our approach supports both fine-grained and coarse-grained scoping, allowing analysts to define scopes for local subprocesses and aggregate them into higher-level processes. For example, analysts may annotate segments of the event log with scopes $P_1$ and $P_2$, and subsequently define a broader scope $P_3$ that encompasses both, thereby capturing a higher level of abstraction.

This capability reflects the hierarchical nature of organizational process landscapes, where operational staff focus on local tasks while managers evaluate performance at the strategic level. By explicitly linking low-level and high-level scopes, analysts can traverse between different levels of abstraction. This not only facilitates drill-down and roll-up style analysis but also supports the exploration of value chains, process handovers, and inter-process dependencies.\footnote{\scriptsize Note that most of the concepts discussed in this paper are the subject of U.S. Patent Application No. 18/944,997}

To operationalize our approach, we provide two supporting toolsets. The first toolset allows analysts to define and embed process scopes into OCELs in a structured and reproducible manner. The second toolset enables the discovery of inter-process interactions within \approachname OCELs. It supports the discovery of how processes interact through shared or linked object types, enabling analysts to identify dependencies among processes based on these connections. 
Together, these tools enhance the analytical power of object-centric process mining by combining expert-driven scoping with scalable, data-driven insights, and by supporting analysis at multiple levels of abstraction.

The remainder of this paper is structured as follows. Section~\ref{sec:Background} provides a brief background. Section~\ref{sec:Approach} introduces both the formal data model and a grammar for specifying scope rules. Section~\ref{sec:Implementation} describes the implementation of the supporting toolsets. Section~\ref{sec:Evaluation} demonstrates the applicability of our approach using a publicly available OCEL log. Finally, Section~\ref{sec:Conclusion} summarizes the main contributions and outlines directions for future research.
\section{Background}\label{sec:Background}

Business Process Management (BPM) has traditionally focused on improving organizational efficiency and effectiveness by analyzing, modeling, and optimizing processes~\cite{weske2012business}. A foundational activity in BPM is defining the scope of a process, which refers to determining what constitutes a process instance and where its boundaries lie~\cite{dumas2018fundamentals}. In process mining, this is typically achieved by selecting a case identifier that guides the extraction of event logs from information systems. As a result, most case-based event logs reflect a single process perspective and omit any representation of inter-process interactions or overlapping responsibilities~\cite{van2020academic}.

In practice, however, processes often do not operate in isolation. They interact, overlap, and co-exist across organizational units, functional domains, and system boundaries. In~\cite{van1999process}, van der Aalst outlined several types of process interoperability, such as chained execution, subcontracting, case transfer, capacity sharing, and loosely coupled execution. These patterns highlight the interconnected nature of modern operations and the need for analytical methods capable of representing and exploring such complexity.

While the need to study these interactions was identified early in process mining research~\cite{van2011intra}, traditional techniques continue to rely on a single case notion that limits the ability to analyze process relationships that span across object types or organizational functions.
The Object-Centric Event Log (OCEL)~\cite{berti2023ocelspecification} was introduced to address single-case process analysis limitations. OCEL supports associating events with multiple related objects, which enables capturing multi-perspective behavior within a single log. This structure has advanced the development of Object-Centric Process Mining (OCPM), which provides tools to analyze processes from various object viewpoints and uncover interactions, dependencies, and parallel execution patterns\cite{aalst2023twin}.

As the richness of OCEL increases, so does the complexity of the analysis~\cite{khayatbashi2025ai}. To address this, recent work has introduced formal operations for adjusting the level of detail in OCEL. These include drill-down, roll-up, unfold, and fold, which allow analysts to transform logs by grouping or separating object types and event semantics~\cite{khayatbashi2025olap,khayatbashi2024advancing}. These operations help adapt the analysis to stakeholder needs, improve model interpretability, and allow switching between fine-grained and high-level perspectives.
However, such techniques are primarily data-centric. They modify the structure of the event log to control the level of abstraction but do not offer an explicit mechanism for defining what should be considered as a process. From a BPM standpoint, abstraction involves not only viewing data at different resolutions but also organizing and understanding how multiple processes relate, overlap, or compose a broader process landscape.

In this paper, we extend OCEL by introducing support for explicitly defined process scopes. This allows analysts to embed structured process definitions into the log based on domain knowledge or analytical objectives. Our approach complements prior work on log transformation by enabling not only more flexible analysis, but also a clearer representation of multiple coexisting processes and their relationships. It supports aggregating behavior across process scopes and enables reasoning at multiple levels of abstraction, from local activities to strategic process structures.
\section{Approach}\label{sec:Approach}

To address the lack of explicit process scope definition in object-centric process mining, we propose a structured extension of the OCEL format that allows analysts to define and embed process boundaries directly within the event log. This enhancement enables processes to be treated as first-class entities in OCEL, making their structure and interrelations explicit.

Our approach is grounded in the observation that process definitions in real-world environments are inherently context-dependent and shaped by domain-specific knowledge. As discussed in the background, this subjectivity poses challenges for automated or system-defined scope discovery. To overcome this, we introduce the notion of \textit{\approachname OCELs}, which incorporates analyst-defined process scopes into the log using a formal specification language and an associated filtering mechanism. This enrichment renders process boundaries transparent, reproducible, and machine-interpretable, thereby enabling structured analysis of coexisting processes and their interactions.

\subsection{Formal definitions}

To formalize our approach, we begin by defining the foundational elements of the OCEL data model. Our formulation adheres to the OCEL 2.0 specification~\cite{berti2023ocelspecification}, while introducing targeted extensions to support the representation of explicit process scopes. These extensions provide the structural basis for embedding scope definitions within the event log and for enabling scoped analysis over object-centric event data.

\begin{definition}\label{def:univ}\normalfont 
    We define the following \textbf{universes}~\cite{berti2023ocelspecification,khayatbashi2024transforming}:\\
     $\univ{eid}$ is the universe of event identifiers, 
        $\univ{\mathit{etype}}$ is the universe of event types,
        $\univ{oid}$ is the universe of object identifiers, 
        $\univ{\mathit{qual}}$ is the universe of qualifiers,
        $\univ{\mathit{time}}$ is the universe of timestamps, 
        $\univ{\mathit{val}}$ is the universe of attribute values,
        $\univ{\mathit{att}}$ is the universe of attribute names, and
        $\univ{\mathit{otype}}$ is the universe of object types, with $\mathit{process}\in\univ{\mathit{otype}}$.
\\We assume that $\univ{\mathit{time}}$ has a smallest element $0$ and a largest element $\infty$.
We also define an object type $\mathit{process}$ to distinguish objects with the type of process, where the object identifier can refer to the name of the process.
\end{definition}

The OCEL structure itself is then defined as a tuple that includes events, objects, their associated attributes, and their qualified relations. This formalism provides the foundation upon which we later build the notion of enriched process scopes.

\begin{definition}\label{def:ocel}
\normalfont
An \textbf{Object-Centric Event Log~(OCEL)}~$L$ is a tuple~$(E,$  $O,\mathit{EA},$ $\mathit{OA},$ $\mathit{evtype},\mathit{evid},\mathit{time},\mathit{objtype},\mathit{objid},\mathit{eatype},\mathit{oatype},\mathit{eaval},
\mathit{oaval},\mathit{E2O},\mathit{O2O})$~\cite{berti2023ocelspecification,khayatbashi2024transforming}, where:
 \begin{itemize}
    \item $E$ and $O$ are sets of events and sets of objects, where $E\cap O = \emptyset$,
    \item $\mathit{EA} \!\subseteq\! \univ{\mathit{att}}$ and $\mathit{OA} \!\subseteq\! \univ{\mathit{att}}$ are sets of
    		attributes for events and objects, respectively,
    \item $\mathit{evtype}: E \rightarrow \univ{\mathit{etype}}$ is a function assigning event types to events,
    \item $\mathit{evid}: E \rightarrow \univ{\mathit{eid}}$ is a function assigning event identifier to events,
    \item $\mathit{evtime}: E \rightarrow \univ{\mathit{time}}$ is a function assigning timestamps to events, 
    \item $\mathit{objtype}: O \rightarrow \univ{\mathit{otype}}$ is a function assigning object types to objects, 
    \item $\mathit{objid}: O \rightarrow \univ{\mathit{oid}}$ is a function assigning object identifier to objects,
    \item $\mathit{eatype}: \mathit{EA} \rightarrow \univ{\mathit{etype}}$ is a function assigning event types to event attributes, 
    \item $\mathit{oatype}: \mathit{OA} \rightarrow \univ{\mathit{otype}}$ is a function assigning object types to object attributes, 
    \item $\mathit{eaval}: (E \times \mathit{EA}) \not\rightarrow \univ{\mathit{val}}$ is a partial function assigning values to (some) event attributes such that $evtype(e) = eatype(ea)$ for all $(e,ea)\in dom(eaval)$,

    \item $oaval: (O \times \mathit{OA} \times \univ{\mathit{time}}) \not\rightarrow \univ{\mathit{val}}$ assigns values to object attributes such that $objtype(o)=oatype(oa)$ for all $(o,oa,t)\in dom(oaval)$,

    \item $\mathit{E2O}\subseteq E \times \univ{\mathit{qual}} \times O$ are the qualified event-to-object relations, and
    
    \item $\mathit{O2O}\subseteq O \times \univ{\mathit{qual}} \times O$ are the qualified object-to-object relations.
 \end{itemize}
\end{definition}

For any partial function $f: D\not\rightarrow R$, if there exists $d\in D$ where $d\notin dom(f)$, we say $f(d)=\perp$. 
We gave an excerpt of the OCEL 2.0 definition that we needed in this paper. We refer the readers to~\cite{berti2023ocelspecification} for the full specification.

To introduce process scoping explicitly, we define a \textit{\approachname OCEL} (denoted POCEL) as an OCEL that contains objects of type \texttt{process}, each representing a specific process scope.
\begin{definition}\label{def:intfun}
    \normalfont
    A \textbf{\Approachname OCEL (POCEL)} is a specific type of OCEL $(E, O,\mathit{EA},$ $\mathit{OA},$ $\mathit{evtype},\mathit{evid},\mathit{time},\mathit{objtype},$ $\mathit{objid},$ $\mathit{eatype},$ $\mathit{oatype},$ $\mathit{eaval},$ $
    \mathit{oaval},\mathit{E2O},\mathit{O2O})$ 
    , where there exists 
    $o,o^\prime\in O,\ e\in E,\ q,q^\prime\in\univ{\mathit{qual}}$
    such that 
    $\mathit{objtype}(o)=\mathit{process}$ and
    $(e, q, o)\ \in\ \mathit{E2O}$
    and 
    $o \neq o^\prime$
    and
    ($o, q^\prime, o^\prime) \in \mathit{O2O}$
    .    
\end{definition}

\subsection{The enrichment language}

To enable the construction of \approachname OCELs in a repeatable and domain-sensitive manner, we introduce a formal specification mechanism, referred to as the \emph{Enrichment language}. This grammar-based language allows analysts to define process scopes by formulating inclusion and exclusion criteria over event and object attributes, object types, and event types. The language is designed to balance expressive power with ease of use, supporting the specification of complex scoping logic through structured and interpretable rule compositions.

The syntax of the language is defined as a context-free grammar. Terminal symbols are drawn from the universe of event and object attributes, object and event type labels, constant values, and logical operators. The grammar supports the construction of filter statements and compound conditions, including conjunctions, disjunctions, and nested expressions, enabling flexible and fine-grained scope definitions.

\begin{definition}[The Enrichment Language Grammar]
    We define the \emph{enrichment language} as a context-free grammar $G = (\Sigma, N, R, S)$, where:
    \begin{itemize}
      \item $\Sigma$ is the set of \emph{terminal symbols}, i.e., $\Sigma = \{\texttt{INCLUDE}, \texttt{EXCLUDE}, \texttt{AND}, \texttt{OR}\} \cup \mathcal{P}(\univ{\mathit{otype}}) \cup \mathcal{P}(\univ{\mathit{etype}}) \cup \univ{\mathit{otype}} \cup \univ{\mathit{etype}} \cup \univ{\mathit{att}} \cup \univ{\mathit{val}}$.
      \item $N$ is the set of \emph{non-terminal symbols}, i.e., $N = \{\langle ruleset \rangle, \langle rule \rangle,$ $\langle statement \rangle,$ $\langle filter\_item \rangle, \langle entity \rangle, \langle attribute \rangle,$ $ \langle operator \rangle,$ $ \langle value \rangle\}.$
      \item $R$ is the set of \emph{production rules} of the form $A ::= \alpha$, defined explicitly by the BNF notation below. Each rule specifies that a non-terminal $A \in N$ can be replaced by a string $\alpha \in (\Sigma \cup N)^*$.
      \item $S = \langle ruleset \rangle$ is the \emph{start symbol}.
    \end{itemize}
    We define the production rules below, where $\epsilon$ denotes the empty string, used to represent unspecified filtering components:
    \begin{align*}
    \langle ruleset \rangle &::= 
    \langle rule \rangle 
    \;|\; (\,\langle ruleset \rangle\, \texttt{AND}\, \langle ruleset \rangle\,)
    \;|\; (\,\langle ruleset \rangle\, \texttt{OR}\, \langle ruleset \rangle\,)
    \\
    \langle rule \rangle &::= 
      \texttt{INCLUDE}\, \langle statement \rangle
      \;|\; \texttt{EXCLUDE}\, \langle statement \rangle\\
      &\;|\; \texttt{INCLUDE}\, \langle statement \rangle
      \, \texttt{AND}\,
      \texttt{EXCLUDE}\, \langle statement \rangle
    \\
    \langle statement \rangle &::= 
    \{\, \langle filter\_item \rangle (\, ,\, \langle filter\_item \rangle )^* \,\}
    \\
    \langle filter\_item \rangle &::= 
    (
    \langle entity \rangle\; ,
    \epsilon ,
    \epsilon ,
    \epsilon )
    \;|\; 
    (
    \langle entity \rangle\; ,
    \langle attribute \rangle\; ,
    \langle operator \rangle\; ,
    \langle value \rangle )
    \\
    \langle entity \rangle &::= 
    \text{an item } \in \univ{\mathit{otype}} \cup \univ{\mathit{etype}} 
    \\
    \langle attribute \rangle &::= 
    \text{an attribute } \in \univ{\mathit{att}}
    \\
    \langle operator \rangle &::= 
    < \;|\; > \;|\; = \;|\; \neq \;|\; \dots
    \\
    \langle value \rangle &::= 
    \text{a value } \in \univ{\mathit{val}}.
    \end{align*}
    Let $L(G)$ be a language generated by this grammar, i.e., $L(G) = \{\, w \in \Sigma^* \mid S \overset{*}{\Longrightarrow} w \}$, where $S \overset{*}{\Longrightarrow} w$ denotes that $w$ can be derived from the start symbol $S$ by applying the production rules in $R$ zero or more times. We call a query $q$ valid if and only if $q \in L(G)$.
\end{definition}

The list of operators can be extended beyond what we defined here to enable richer filtering. A ruleset written in this language defines the conditions under which events and objects should be included in or excluded from the scope of a process. A query is considered valid if and only if it belongs to the language defined by the grammar.

\subsection{The enrichment procedure}
Given a valid ruleset expressed in the the enrichment language, we apply a transformation procedure to generate an \approachname OCEL. This procedure evaluates the ruleset over the original log, filters relevant events and associated objects, and instantiates a new process object representing the defined scope. The newly created process object is assigned a unique identifier and linked to all selected entities through appropriate OCEL relations. The result is a \approachname OCEL in which the process scope is explicitly represented and structurally embedded in compliance with the OCEL schema.

We do not formalize the detailed operational semantics of the enrichment process, as the transformation is deterministic and straightforward once the ruleset is defined. Instead, the remainder of the paper focuses on demonstrating how \approachname OCELs enable systematic discovery and visualization of inter-process interoperability, supported by tools and artifacts implemented around the proposed language and its semantics - as a proof of concept.

\section{Implementation}\label{sec:Implementation}

We developed two complementary tools to support the proposed approach. The first, \textbf{Procellar}, enables analysts to define rulesets based on which they export an \approachname OCEL from an existing OCEL. The second, \textbf{Business Execution Graph}, facilitates the analysis of process interactions based on \approachname OCEL. Both tools feature front-end and back-end components, implemented in JavaScript and Python, respectively.

\subsection{Procellar}
Procellar allows users to import an OCEL log and define process scope rulesets through an interactive interface. Each ruleset consists of one or more rules, combined using logical operators as defined by the grammar of the enrichment language. Once defined, rulesets can be applied to enrich the OCEL, resulting in a \approachname version of the log. Both the \approachname OCEL and the corresponding rulesets can be exported for further analysis or reuse.

The tool provides two modes for rule definition: a \emph{basic mode} and an \emph{advanced mode}. In basic mode, users define rules by including or excluding specific object types without filtering attribute values. This mode simplifies the scoping process when fine-grained filtering is unnecessary. In contrast, advanced mode allows the specification of complex scoping criteria involving combinations of event or object attributes.

\begin{figure}[t!]
    \centering
    \includegraphics[width=1\linewidth]{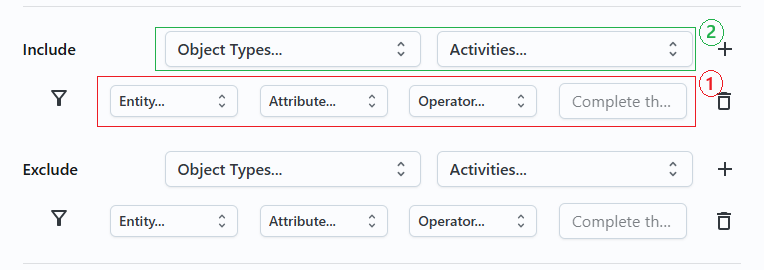}
    \caption{Rule definition for enabling \approachname OCEL with process notion in \textbf{Procellar}, where area marked with (1) represents a filter item, and area marked with (2) represents implementation adjustment for better usability for including object types or event types without any filter.}
    \label{fig:ruledefinition}
	\vspace{-1.2\baselineskip}    
\end{figure}

\figurename~\ref{fig:ruledefinition} illustrates the rule definition interface in advanced mode. The area marked by (1) shows the rule construction panel, where users can select the entity type (object or event), choose attributes, specify operators, and enter corresponding values. The area marked by (2) provides a shortcut for including object and event types, i.e., activities, without attribute filtering, which automatically generates rules with empty filtering conditions as defined in the grammar.
Procellar is available as an open-source project on GitHub\footnote{\scriptsize \url{https://github.com/hudsonjychen/procellar}}.

\subsection{Business Execution Graph (BEG)}
The Business Execution Graph (BEG) tool enables the analysis and visualization of interactions among processes in a \approachname OCEL. It supports the identification of shared objects, dependencies, and potential coordination structures between processes.

Users can configure the visualization through various settings. Node sizes can be adjusted based on object counts or object type diversity. Edge labels can be customized to display object types, total shared objects, or average flow times, enabling analysts to explore process dependencies considering different measures. For example, labeling edges by object type allows investigation of how different object types are involved in process interactions.

In addition, node colors can be dynamically styled based on the number of incoming, outgoing, or total edges. The network visualization can be exported in PNG format for reporting or presentation purposes. The tool also supports exporting the graph to the VOSviewer format for advanced network analysis and clustering.
Business Execution Graph is also available as an open-source tool on GitHub\footnote{\scriptsize \url{https://github.com/hudsonjychen/business-execution-graph}}.

\section{Demonstration}\label{sec:Evaluation}

To demonstrate the applicability and flexibility of our approach, we applied the proposed toolsets to a publicly available Object-Centric Event Log (OCEL) from the logistics domain\footnote{\scriptsize \url{https://ocel-standard.org/event-logs/simulations/logistics/}}.
This event log captures an end-to-end logistics scenario, beginning with customer order placement and concluding with the transportation of goods via various long-distance carriers, such as ships and trains. It involves interactions among multiple object types, including orders, goods, transport documents, and containers. We applied our enrichment methodology by defining four distinct but interrelated processes (through 4 defined rules), each representing a key phase in the logistics value chain:

\begin{itemize}
    \item \textit{Order Management}: Includes events related to the creation and management of customer orders, concluding with the issue of transport documents needed for starting the goods management process (594 events).
    \item \textit{Goods Management}: Covers preparatory activities, including ordering empty containers, collecting goods, and loading goods into trucks, making them ready for dispatch to terminals (13155 events).
    \item \textit{Transportation Management}: Focuses on the movement of containers using ground vehicles between warehouses and terminals, capturing logistical transport operations before long-distance carriers' activities (18314 events).
    \item \textit{Export Management}: Represents the final phase, where containers are scheduled and loaded onto ships or trains for long-distance export (2132).
\end{itemize}

\begin{figure}[t!]
    \centering
    \includegraphics[width=1\linewidth]{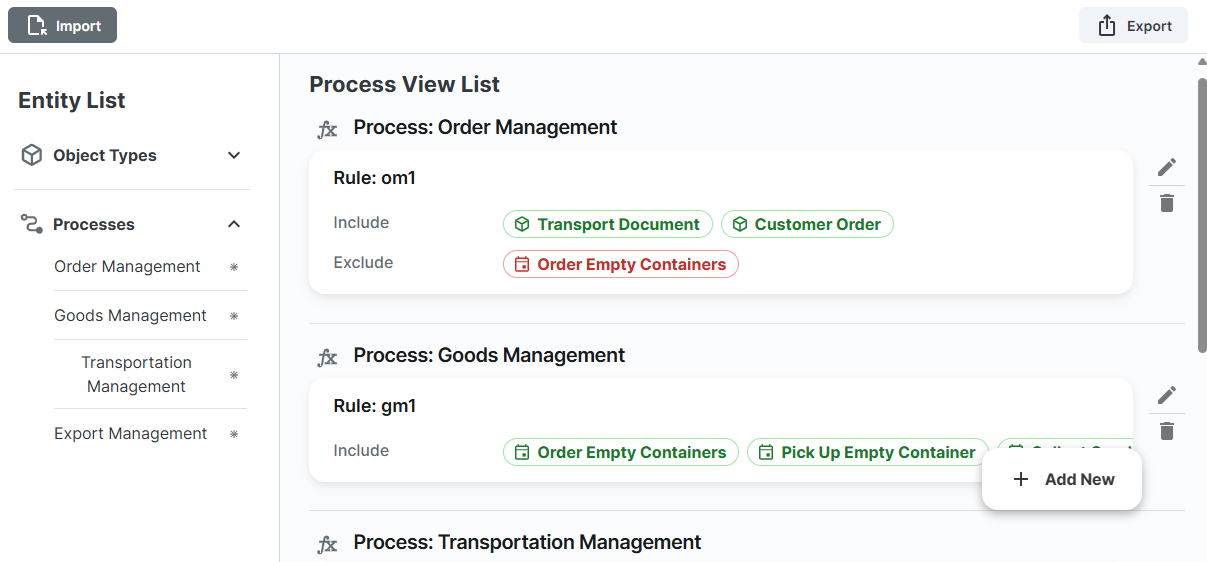}
    \caption{Rule configuration in \textbf{Procellar} for projecting logistics processes}
    \label{fig:case-rules}
	\vspace{-1.2\baselineskip}    
\end{figure}

\figurename~\ref{fig:case-rules} shows a screenshot from the Procellar tool, illustrating how the process scopes were defined using a combination of object types and event filtering. Each process was specified through a corresponding ruleset, allowing the log to be modularly partitioned and enriched. The resulting \approachname  OCEL and the projection rulesets are publicly available in our GitHub repository, enabling reproducibility and further analysis by the research community\footnote{\scriptsize \url{https://github.com/shahrzadkhayatbashi/Process-Level-OCPM}}.

\begin{figure}[t!]
    \centering
    \includegraphics[width=1\linewidth]{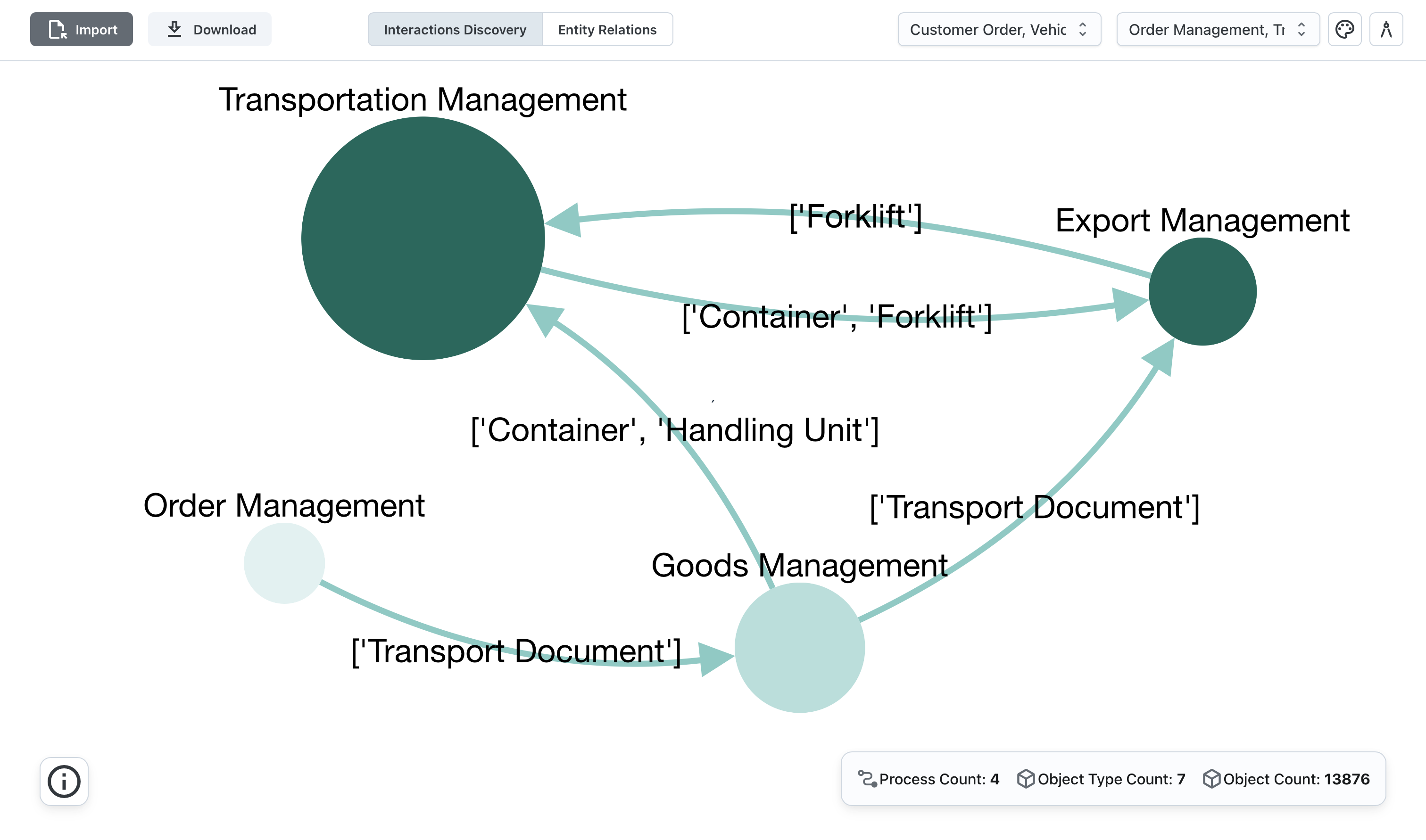}
    \caption{\textbf{BEG} visualization of interactions among logistics processes}
    \label{fig:BEG}
	\vspace{-1.2\baselineskip}    
\end{figure}

The \approachname  OCEL served as input to the Business Execution Graph (BEG), which was used to visualize interactions and dependencies between the defined processes. \figurename~\ref{fig:BEG} presents the resulting graph, where each node represents a process and edges capture object types that are handed over between processes. Node sizes are proportional to the number of associated object instances, while edge labels denote the object types that facilitate transitions. Node colors indicate the number of incoming edges, providing visual clues about process centrality and interconnectivity.

As the figure illustrates, the \textit{Order Management} process initiates the chain by transmitting \textit{Transport Document} objects to the \textit{Goods Management} process. The latter, in turn, provides inputs for both the \textit{Transportation Management} and \textit{Export Management} processes. Notably, the execution of \textit{Transportation Management} relies on coordination involving both \textit{Container} and \textit{Handling Unit} object types. The \textit{Export Management} process depends on the outcomes of \textit{Transportation Management}, especially in the use of \textit{Forklifts} to transfer containers to export terminals. A notable pattern observed in the graph is a loop, indicating forklifts returning empty after container delivery.

This demonstration validates the effectiveness of our approach in handling object-centric event data. By enabling structured projection into modular process scopes and providing visual insights into process interactions, our toolset supports a systematic analysis of complex, interconnected operations. The resulting \approachname  OCEL and interaction graph can facilitate a deeper understanding of process dependencies, handovers, and coordination mechanisms within object-rich domains.

\section{Conclusion}\label{sec:Conclusion}

This paper addresses a key limitation in Object-Centric Process Mining (OCPM), namely the absence of explicit process scope representations in current object-centric event data standards such as OCEL 2.0. To overcome this challenge, we proposed a \approachname framework that allows analysts to define and embed process scopes directly within OCEL. Our contribution includes a formal grammar-based language for specifying scoping rules and two open-source tools (Procellar and Business Execution Graph) that operationalize this approach.

By enabling the systematic construction of \approachname OCELs, our framework bridges the gap between data-driven analysis and domain-specific process knowledge. The introduced tools support modular process projection, scope-aware event filtering, and the visualization of inter-process dependencies based on object-centric interactions. Our demonstration using a public logistics dataset illustrates how the approach facilitates the discovery of process boundaries, overlaps, and coordination patterns in complex operational environments.
The proposed method enhances the expressiveness and interpretability of OCPM, offering analysts greater control over the level of abstraction and enabling structured reasoning about process architecture. It supports both fine-grained analysis and high-level process aggregation, making it suitable for a wide range of analytical and organizational needs.

Future work may focus on extending the language to support temporal constraints and behavioral patterns within scopes, exploring semi-automatic discovery approaches to support analyst-defined rules, and integrating the approach into enterprise modeling environments to support strategic alignment and process governance. In addition, we aim to apply this approach in a real case, where we can evaluate how a \approachname approach would be perceived by analysts, and can eventually improve the overall Object-Centric Process Mining experience and adaptation in practice.

\bibliographystyle{unsrt} 
\bibliography{references}

\end{document}